\documentclass[twocolumn,floatfix,prc,showpacs,aps,superscriptaddress]{revtex4-1}
\usepackage{graphicx,bm}
\usepackage{amssymb,amsmath}
\usepackage[colorlinks,citecolor=blue]{hyperref}
\begin{document}


\title{On the nuclear robustness of the r process in neutron-star
  mergers}

\author{Joel de Jes\'us Mendoza-Temis} 
\affiliation{Institut f{\"u}r Kernphysik
  (Theoriezentrum), Technische Universit{\"a}t Darmstadt,
  Schlossgartenstra{\ss}e 2, 64289 Darmstadt, Germany}

\author{Meng-Ru Wu} 
\affiliation{Institut f{\"u}r Kernphysik
  (Theoriezentrum), Technische Universit{\"a}t Darmstadt,
  Schlossgartenstra{\ss}e 2, 64289 Darmstadt, Germany}

\author{Karlheinz Langanke}
\affiliation{GSI Helmholtzzentrum f\"ur Schwerionenforschung,
  Planckstra{\ss}e~1, 64291 Darmstadt, Germany}
\affiliation{Institut f{\"u}r Kernphysik
  (Theoriezentrum), Technische Universit{\"a}t Darmstadt,
  Schlossgartenstra{\ss}e 2, 64289 Darmstadt, Germany}

\author{Gabriel Mart\'{\i}nez-Pinedo} 
\affiliation{Institut f{\"u}r Kernphysik
  (Theoriezentrum), Technische Universit{\"a}t Darmstadt,
  Schlossgartenstra{\ss}e 2, 64289 Darmstadt, Germany}
\affiliation{GSI Helmholtzzentrum f\"ur Schwerionenforschung,
  Planckstra{\ss}e~1, 64291 Darmstadt, Germany}

\author{Andreas Bauswein} 
\affiliation{Department of Physics, Aristotle
  University of Thessaloniki, 54124 Thessaloniki, Greece}
\affiliation{Heidelberger Institut f\"ur Theoretische Studien,
  Schloss-Wolfsbrunnenweg~35, 69118~Heidelberg, Germany}

\author{Hans-Thomas Janka}
\affiliation{Max-Planck-Institut f\"ur Astrophysik, Postfach 1317,
  85741 Garching, Germany}

\begin{abstract}
  We have performed r-process calculations for matter ejected
  dynamically in neutron star mergers based on a complete set of
  trajectories from a three-dimensional relativistic smoothed particle
  hydrodynamic simulation with a total ejected mass of
  $\sim 1.7\times 10^{-3}$~M$_\odot$. Our calculations consider an
  extended nuclear network, including spontaneous, $\beta$- and
  neutron-induced fission and adopting fission yield distributions
  from the ABLA code. In particular we have studied the sensitivity of
  the r-process abundances to nuclear masses by using different
  models. Most of the trajectories, corresponding to 90\% of the
  ejected mass, follow a relatively slow expansion allowing for all
  neutrons to be captured. The resulting abundances are very similar
  to each other and reproduce the general features of the observed
  r-process abundance (the second and third peaks, the rare-earth peak
  and the lead peak) for all mass models as they are mainly determined
  by the fission yields. We find distinct differences in the
  predictions of the mass models at and just above the third peak,
  which can be traced back to different predictions of neutron
  separation energies for r-process nuclei around neutron number
  $N=130$. In all simulations, we find that the second peak around
  $A\sim 130$ is produced by the fission yields of the material that
  piles up in nuclei with $A\gtrsim 250$ due to the substantially
  longer beta-decay half-lives found in this region. The third peak
  around $A \sim 195$ is generated in a competition between neutron
  captures and $\beta$ decays during r-process freeze-out.  The
  remaining trajectories, which contribute 10\% by mass to the total
  integrated abundances, follow such a fast expansion that the r
  process does not use all the neutrons. This also leads to a larger
  variation of abundances among trajectories as fission does not
  dominate the r-process dynamics.  The resulting abundances are in
  between those associated to the r and s processes. The total
  integrated abundances are dominated by contributions from the slow
  abundances and hence reproduce the general features of the observed
  r-process abundances.  We find that at timescales of weeks relevant
  for kilonova light curve calculations, the abundance of actinides is
  larger than the one of lanthanides. This means that actinides can be
  even more important than lanthanides to determine the photon
  opacities under kilonova conditions. Moreover, we confirm that the
  amount of unused neutrons may be large enough to give rise to
  another observational signature powered by their decay.
\end{abstract}
\maketitle

\section{Introduction}

The astrophysical r process produces about half of the heavy elements
in the Universe, including all of the
actinides~\cite{Burbidge.Burbidge.ea:1957,Cameron:1957}. It is
commonly accepted that it occurs as a sequence of neutron captures and
$\beta$ decays in environments with extreme neutron densities. Under
such conditions neutron captures are much faster than $\beta$ decays
and the r-process path runs through nuclei with large neutron excess
far off stability
\cite{Cowan.Thielemann.Truran:1991,arnould.goriely.takahashi:2007}. 

The actual astrophysical site of the r process is yet not known. For
many years the neutrino-driven wind from the surface of a freshly born
neutron star in a core-collapse supernova has been the favored
site~\cite{Woosley.Wilson.ea:1994}. However, recent supernova
simulations with improved nuclear input, realistic neutrino transport
and advanced multi-dimensional treatment of
hydrodynamics~\cite{Janka:2012} indicate that the conditions of the
matter ejected in the wind (entropy, expansion velocity,
proton-to-neutron ratio) are not suited to support an r process which
produces the elements in the third peak (around mass number $A \sim
195$) and beyond.  However, the neutrino-driven wind might
significantly contribute to the abundance of the lighter elements up
to the second peak ($A \sim 130$)~\cite{Martinez-Pinedo.Fischer.ea:2012,Roberts.Reddy.Shen:2012,Martinez-Pinedo.Fischer.Huther:2014}. The shortfall of the
neutrino-driven wind model to produce the heavier r-process elements
has revived the interest in another potential site, the merger of two
neutron stars (NS merger)~\cite{Lattimer.Schramm:1974,Freiburghaus.Rosswog.Thielemann:1999}.

Simulations of NS mergers indicate that the matter ejected during the
dynamical phase is very neutron rich with extremely large
neutron-to-seed ratios
($R_{n/s} >
400)$~\cite{Goriely.Bauswein.Janka:2011,Korobkin.Rosswog.ea:2012,Bauswein.Goriely.Janka:2013};
i.e.\  there are many neutrons which can be captured by seed nuclei
transporting matter to very heavy nuclei in the region of the nuclear
chart where decay by fission is possible. The intermediate-mass
fission yields are then subject to neutron captures establishing the
occurrence of a few fission cycles, which are expected to produce the
heavier r-process elements in a rather robust way with nearly relative
solar abundances. We note that such a robust scenario is quite
attractive as it might explain the occurrence of r-process elements
above the second peak in solar proportion as observed in very old
metal-poor stars~\cite{Sneden.Cowan.Gallino:2008}. Recent simulations
suggest that the electron fraction of part of the ejecta may be raised
by neutrino
processes~\cite{Wanajo.Sekiguchi.ea:2014,Sekiguchi.Kiuchi.ea:2015,Goriely.Bauswein.ea:2015}. However,
the sensitivity of these results to the treatment of neutrino
radiation transport in NS mergers simulations remains to be
explored. In this study, we work under the assumption that all ejecta
remain neutron rich.

An additional source of ejected material with relevance for r-process
nucleosynthesis comes from the accretion disk formed around the
compact object resulting from the merger. However, the conditions in
these ejecta are more sensitive to the details of the astrophysical
parameters and microphysics
included~\cite{Fernandez.Metzger:2013,Just.Bauswein.ea:2015,Metzger.Fernandez:2014,Perego.Rosswog.ea:2014}. Here
we will consider only the initial dynamical ejecta.

To serve as the site for the production of heavy r-process elements
and to explain the observation of the elemental abundances in solar
proportion in metal-poor stars, the r process in NS mergers should
not depend on particular astrophysical conditions, e.g.\ on the
specific combination of NS masses in the merging binary system.
It is unsatisfactory for r-process abundance simulations that most
nuclei encountered during the process have yet not been produced in
the laboratory and hence their properties depend on nuclear models and
are yet quite uncertain.

The sensitivity of r-process nucleosynthesis in dynamical ejecta of NS
mergers to the astrophysical conditions, i.e.\ neutron star masses,
orbit parameters, has been studied
in~\cite{Goriely.Bauswein.Janka:2011,Korobkin.Rosswog.ea:2012,Bauswein.Goriely.Janka:2013,Sekiguchi.Kiuchi.ea:2015}.
Korobkin \emph{et al.} have performed r-process simulations for a set
of NS mergers consisting of various combinations of neutron stars in
the relevant mass range between 1 and
2~M$_\odot$~\cite{Korobkin.Rosswog.ea:2012}, while Bauswein \emph{et
  al.} have explored the influence of various equations of state on
the merger dynamics and
nucleosynthesis~\cite{Bauswein.Goriely.Janka:2013} (see also
ref.~\cite{Sekiguchi.Kiuchi.ea:2015}). It turns out that the specific
treatment of the merger dynamics, e.g. Newtonian vs General
Relativistic mechanics, leads to fundamentally different mass ejection
dynamics~\cite{Korobkin.Rosswog.ea:2012,Bauswein.Goriely.Janka:2013}.
Importantly for our discussion both groups find, within their
treatment, nearly identical abundance distributions between the second
and third peaks for all of the 23 combinations of neutron
stars~\cite{Korobkin.Rosswog.ea:2012} or the various adopted equations
of state~\cite{Bauswein.Goriely.Janka:2013}, pointing to virtually no
sensitivity of the relative abundance of heavy elements on the
astrophysical conditions of the mergers. However, both groups also
found strong sensitivity of the abundances to the treatment of
fission. Indeed, the calculations of~\cite{Korobkin.Rosswog.ea:2012}
show an abundance hole around $A\sim 140$, relative to solar, while in
the study of ref.~\cite{Bauswein.Goriely.Janka:2013} the second peak
is noticeably shifted to larger mass numbers.  Both effects are
related to the treatment of fission adopted in the respective
simulations.

This dependence on fission as well as the effect of half lives and
neutron separation energies (masses) on the r-process abundances in NS
merger simulations has been the focus of three simultaneous and
independent studies. Eichler \emph{et al.} have confirmed the strong
sensitivity on the fission yield
distribution~\cite{Eichler.Arcones.ea:2015}.  However, they succeeded
to show that fission yields derived with the ablation-abrasion model
code ABLA~\cite{Gaimard.Schmidt:1991,Kelic.Valentina.Schmidt:2009},
which is based on the statistical model, cured the shortcomings in the
abundance distribution above the second peak. The ABLA code is
adjusted to reproduce fission data and considers the evaporation of
free neutrons before and after the fission process. However, these
neutrons as well as those which are produced by the decay of the
fission fragments can be captured by nuclei after freeze-out resulting
in a slight shift of the third peak to heavier mass numbers than is
observed in the solar abundances. Eichler {\it et
  al.}~\cite{Eichler.Arcones.ea:2015} argue that such a shift can be
avoided by faster $\beta$ decays than those predicted by the Finite
Range Droplet Model (FRDM) model which has been adopted in their
study. In an independent study Caballero \emph{et al.} come to the
same conclusion~\cite{Caballero.Arcones.ea:2014} when they compare
results of NS merger r-process simulations performed with the half
lives based on the FRDM model with those obtained by replacing these
half lives with faster values derived by QRPA calculations on top of
the energy density functional (EDF) of
Fayans~\cite{Borzov:2006}. Importantly, these faster EDF-based half
lives agree well with recent experimental data obtained for nuclei
close and on the r-process path, including data for neutron-rich
nuclei towards the $N=126$ waiting
points~\cite{Kurtukian-Nieto.Benlliure.ea:2014}. Here, the EDF-based
half-lives, in close agreement with recent large-scale shell model
calculations~\cite{Zhi.Caurier.ea:2013,Suzuki.Yoshida.ea:2012}, point
to the importance of forbidden transitions to the $\beta$ decays.

In this manuscript, we study the effect of nuclear masses and neutron
capture rates on the r process in NS mergers. Masses are particularly
important as they, via the neutron separation energies, define the
r-process path in the nuclear chart and secondly they are crucial
ingredients in the statistical model calculation of neutron capture
cross sections. In the following we will derive neutron separation
energies as well as neutron capture cross sections consistently from
the same mass models. Furthermore photodissociation cross sections are
obtained by detailed balance from the capture cross sections.  As the
masses of the extremely neutron-rich nuclei on the r-process path are
not known experimentally, they have to be modeled. For many years,
masses derived on the basis of the Finite Range Droplet Model
(FRDM)~\cite{Moeller.Nix.ea:1995} and the ETFSI model (Extended Thomas
Fermi Model with Strutinski
Integral~\cite{Pearson.Nayak.Goriely:1996}) have been a standard in
r-process simulations. Recently Wang and Liu developed an alternative
microscopic-macroscopic mass model (Weizs\"acker-Skyrme or WS3
model~\cite{Liu.Wang.ea:2011}), which employs a Skyrme energy density
formula as a macroscopic basis, which is then microscopically
supplemented by shell corrections.  Guided by intuition derived from
the interacting shell model, Duflo and Zuker developed a mass formula
based on a systematic description of occupation numbers and taking
special care of the role of intruder
states~\cite{Duflo.Zuker:1995}. In this work we will use the
Duflo-Zuker mass formula with 31 parameters (DZ31). Finally advances
in computing resources make it now possible to derive mass formulas on
the basis of microscopic nuclear many-body models, like the
Hartree-Fock-Bogoliubov (HFB) model. In a sequence of continuous
improvements of the Skyrme functional, which is the basis of their
model, Goriely and collaborators have succeeded to obtain an HFB mass
model which is comparative to the phenomenological models, like FRDM
and ETFSI, and can be globally applied in r-process
simulations~\cite{Goriely.Chamel.Pearson:2009}.  In our calculations,
we will use the model HFB21~\cite{Goriely.Chamel.Pearson:2010}.

The sensitivity of the r-process abundances in NS mergers to
variations of nuclear masses has, so far, been insufficiently
explored. There are calculations in the literature based on different
mass models but it is often difficult to compare the results due to a
combination of factors including rather different assumptions about
the astrophysical conditions, different approaches for the calculation
of the relevant neutron capture rates, and/or the use of very
different fission yield distributions. In this work, we aim at
exploring how robust the r-process abundances are against variations
of the nuclear masses keeping the astrophysical conditions
unchanged. 
We use four different mass models (FRDM, WS3, DZ31, HFB21)
for the calculation of the neutron capture rates that enter in
the r-process simulations. Furthermore, we aim at determining the
nuclear origin of the robust r-process pattern observed in several
neutron star merger simulations. It is often stated that NS mergers
produce a robust r process due to fission cycling. However, this
statement simply expresses the fact that fission cycling is
unavoidable due to the large neutron-to-seed ratios reached in NS
merger ejecta without really explaining the nuclear mechanism
responsible for the robustness.

We have used the full set of trajectories based on a 3-dimensional
relativistic simulation of the merger of two neutron stars with
1.35~M$_\odot$, which is expected to be the most frequent NS merger
system.  Although these trajectories cover a broad range of dynamical
parameters, including a large variation in neutron-to-seed values, we
will argue that they can be best classified into two categories with
respect to their final nucleosynthesis abundance yields: i) For the
majority of trajectories neutrons are nearly completely depleted at
the end of r-process nucleosynthesis, producing significant abundances
of heavy nuclei in the fissioning region, which, in turn, produce the
$A \sim 130$ peak in the r-process abundance distribution by their
decay yields. The resulting r-process abundances show quite similar
patterns.\ ii) Around 10\%, in mass, of the trajectories follow a very
fast expansion. The associated low matter densities result in the fact
that not all neutrons have been captured at the end of the r
process~\cite{Just.Bauswein.ea:2015,Metzger.Bauswein.ea:2015}.  In
this scenario the abundance of heavy nuclei in the fissioning region
is rather low and the second r-process peak ($A \sim 130$) is a
consequence of stalled matter flow due to the relatively long
half-lives of nuclei in the vicinity of the magic neutron number
$N=82$. In this scenario the final r-process abundance distribution is
rather sensitive to nuclear structure effects.

We stress that our results should not be
considered as a prediction of the typical r-process yields from NS
mergers to be used in chemical evolution studies. This will require a
complete set of trajectories for both the dynamical and disk ejecta as
done in ref.~\cite{Just.Bauswein.ea:2015}.

Our paper is organized as follows. In the next section we give a brief
description of our r-process simulations and the input being used. The
results of our simulations for the r-process abundances and their
dependence on the adopted mass models are presented and discussed in
section~\ref{sec:results}. Finally, we conclude in
section~\ref{sec:conclusions}.

\section{NS merger trajectories and nuclear input}
\label{sec:ns-merg-traj}

The r-process calculations in this work are based on fluid element
trajectories that were extracted from hydrodynamical simulations of NS
mergers to represent the conditions of matter becoming unbound from
such events. The merger simulations were performed with a
three-dimensional relativistic smoothed particle hydrodynamics
code~\cite{Oechslin.Rosswog.Thielemann:2002,Oechslin.Janka.Marek:2007,%
  Bauswein.Janka.Oechslin:2010,Bauswein.Goriely.Janka:2013}, which
imposes conformal flatness on the spatial three-metric to solve the
Einstein equations in an approximate
manner~\cite{Isenberg.Nester:1980,Wilson.Mathews.Marronetti:1996}. The
calculations started from initial data representing close binary
systems in quasi-equilibrium a few orbits before merging. Initially
the NS matter was in neutrino-less beta-equilibrium at zero
temperature. The initial electron fraction was advected with the fluid
during the hydrodynamical simulations without taking into account
neutrino processes. This simplification was a reasonably good
approximation in Newtonian models (e.g.~\cite{Ruffert.Janka.ea:1997}),
but the impact of neutrinos in relativistic merger models requires
further investigation~\cite{Wanajo.Sekiguchi.ea:2014,%
Sekiguchi.Kiuchi.ea:2015,Goriely.Bauswein.ea:2015}. NS matter is
modelled with the TM1 EoS~\cite{Sugahara.Toki:1994,%
Hempel.Fischer.ea:2012,Steiner.Hempel.Fischer:2013}, which leads to
a NS radius of 14.49~km for a 1.35~M$_\odot$ NS and a maximum
gravitational mass of 2.21~M$_\odot$ for non-rotating NSs.

In this work we focused on a binary system of two NSs with
gravitational masses of 1.35~M$_\odot$, which may be representative
for the observed double NS systems (see e.g.~\cite{Lattimer:2012} for
a compilation of measured binary NS masses). In the simulation we
found most unbound matter originating from the contact interface
during the coalescence (see~\cite{Bauswein.Goriely.Janka:2013} for a
detailed description of the merger dynamics and ejection
mechanism). Originating from the inner NS crust these ejecta are very
neutron-rich. The time-step limitations of the NS simulations allowed
to follow the ejecta only up to $t=t_0\sim$ several 10 milliseconds. 
Hence, we have analytically continued the trajectories for $t>t_0$
by evolving the density assuming a 
homologous expansion, i.e. $\rho(t) = \rho(t_0) \left(t_0/t\right)^3$.

We have calculated the r-process abundances for 528 trajectories with
a total mass of $\sim 1.70 \times 10^{-3}$~M$_\odot$.  As we will show
below, the r-process nucleosynthesis dynamics and its final abundance
distribution implies to classify the trajectories with respect to two
competing rates: the depletion rate due to neutron captures on seed
nuclei and the hydrodynamic expansion rate.  The neutron depletion
rate $\lambda_n$ can be approximated by
\begin{equation}
\lambda_n=\frac{d(\ln Y_n)}{dt}\approx\frac{\rho
  Y_s}{m_u}\overline{\langle\sigma v\rangle} 
=\frac{\rho Y_n}{m_u R_{n/s}}\overline{\langle\sigma v\rangle},
\end{equation}
where $Y_{n(s)}$ is the number fraction of neutrons (seed nuclei)
and $\overline{\langle\sigma v\rangle}$ is the neutron
capture rate averaged over the seed nuclei. The 
hydrodynamic expansion rate for homologous expansion can be 
calculated by
\begin{equation}
\lambda_d= - \frac{d(\ln\rho)}{dt}=\frac{3}{t}.
\end{equation}

\begin{figure*}[htb]
  \centering
  \includegraphics[width=\linewidth]{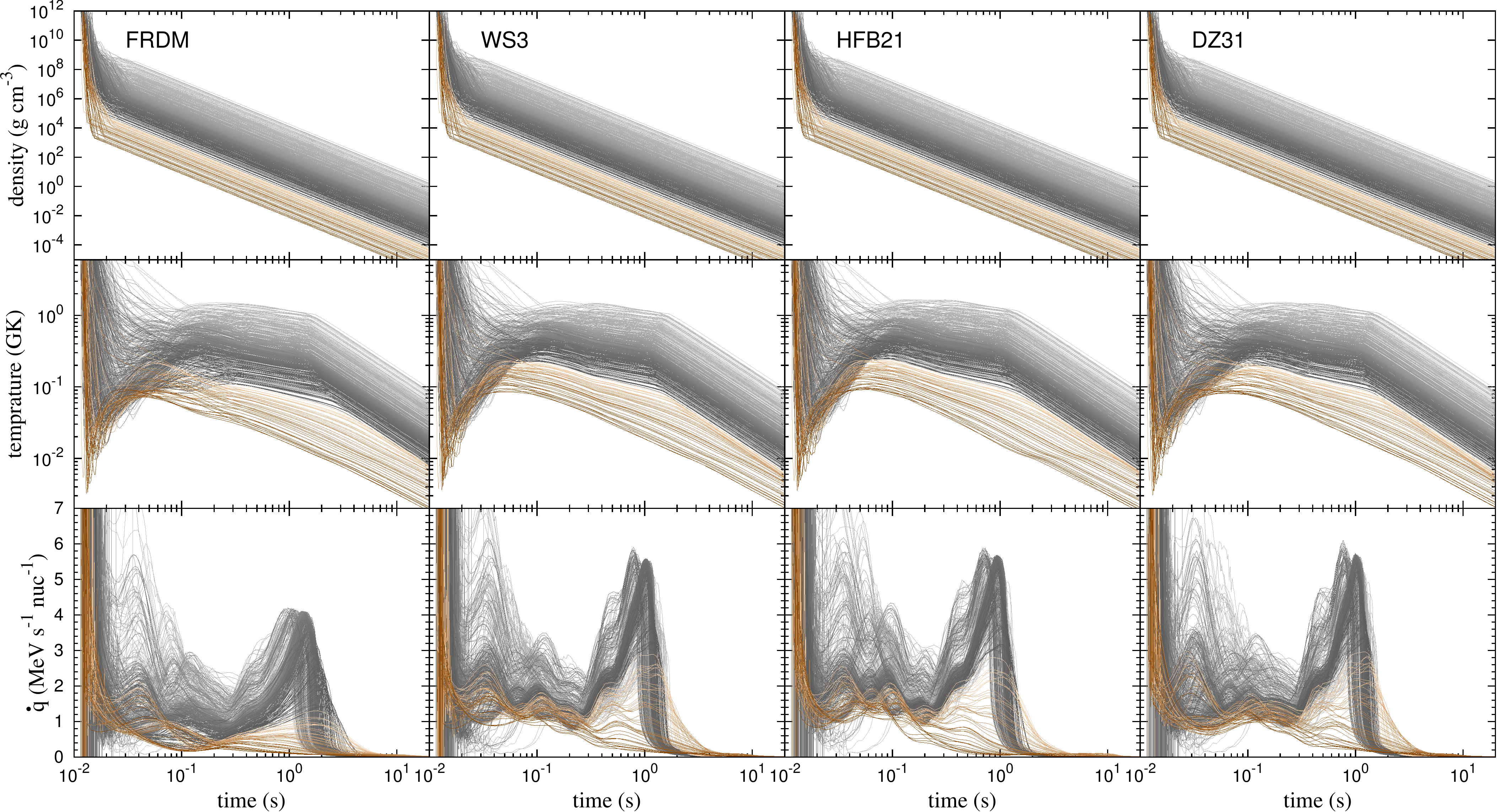}
  \caption{(Color online) Evolution of the different thermodynamical
    variables for all trajectories for different mass models. 
    The gray (brown) curves correspond to the ``slow (fast) ejecta''
    (see text for the definition).
    The upper panels show
    the evolution of density that is identical for all cases. The
    middle panel shows the evolution of temperature that considers the
    feedback due to the energy generation shown in the lower
    panel.\label{fig:reheat}}
\end{figure*}

For most of the ejecta (484 trajectories with a total mass of
$\sim 1.57\times 10^{-3}$~M$_\odot$), nearly all initial neutrons can
be captured as $\lambda_n \gtrsim\lambda_d$ until the end of the r
process at $\sim 1$~s.  This implies that the density should be larger
than a threshold value to allow for the complete capture of neutrons:
$\rho_{\text{th}} \approx 0.5$~g~cm$^{-3}$ at $t\sim 1$~s for typical
values of $Y_n\sim 1$,
$\overline{\langle\sigma v\rangle}\sim 10^{-20}$~cm$^3$~s$^{-1}$, and
$R_{n/s}\sim 10^3$.  We will call these trajectories ``slow ejecta"
in the following. They are labelled by grey curves in
Fig.~\ref{fig:reheat}.  On the other hand, about $\sim 10\%$ of the
ejecta (44 trajectories, $\sim 1.28\times 10^{-4}$~M$_\odot$)
initially expand extremely fast for a few ms reaching density values
lower than the threshold discussed above.  This results in a neutron
depletion rate smaller than the hydrodynamic expansion rate. As a
consequence free neutrons are left at the end of the r process, which
may potentially influence the observed light curves of the ejecta due
to the decay of neutrons~\cite{Metzger.Bauswein.ea:2015}.  In the
following we will call these trajectories ``fast ejecta''. They are
labelled by brown curves in Fig.~\ref{fig:reheat}.

We have started our r-process calculations at temperatures of
$T = 6$~GK, with densities ranging from $\rho\sim 10^7$~g~cm$^{-3}$ to
$\sim 3\times 10^{13}$~g~cm$^{-3}$.  The initial matter compositions
have been determined assuming the matter to be in Nuclear Statistical
Equilibrium (NSE).  The initial neutron-to-seed ratios ($R_{n/s}$) of
those ejecta range from 400 to 2000.

While the details of the initial composition
depend on the conditions of the trajectories and on the mass models
used, we can generally observe that they are given by neutron-rich
nuclei centered around magic neutron numbers $N=50$ and $N=82$, while
proton numbers in the iron-nickel range are favored in the first case
and around strontium-zirconium in the second. The relative weight of
these two composition peaks depends on entropy, favoring the peak
around mass number $A \sim 120$ relative to the one around $A \sim 80$
with decreasing entropy.

Starting from these initial compositions we have followed the
r-process evolution by a large network. The dynamics of the process
was governed by the astrophysical trajectories, however, consistently
corrected for reheating by energy release in nuclear reactions, as we
describe below. Our network includes more than 7300 nuclei which cover
the nuclear chart from free nucleons up to $^{313}$Ds. As nuclear
reactions among these nuclei we considered charge particle reactions,
neutron captures and its inverse process, photo-dissociation, and
$\beta$ and $\alpha$ decay and fission. We have derived the neutron
capture rates consistently for each individual mass model within the
statistical model using the code MOD-Smoker~\cite{Loens:2010}. The
photodissociation rates were obtained from the neutron capture rates
by detailed balance. For nuclei, for which the half lives are not
known experimentally, we have adopted the $\beta$ decay (and $\beta$
delayed neutron emission) rates from the compilation of M\"oller
\emph{et al.}~\cite{Moeller.Pfeiffer.Kratz:2003}, which was derived
from QRPA calculations on top of the FRDM mass model. We used the
parametrization of Ref.~\cite{Dong.Ren:2005} of the Viola-Seaborg
formula to estimate the $\alpha$-decay rates, which become relevant
for heavy nuclei beyond lead. Finally, for nuclei with $Z>83$, where a
competition between $(n,\gamma)$ and neutron induced fission can take
place, we used neutron-induced reaction rates taken
from~\cite{Panov.Korneev.ea:2010} that are based on the FRDM mass
model~\cite{Moeller.Nix.ea:1995} and the Thomas-Fermi fission barriers
of Myers and Swiatecki~\cite{Myers.Swiatecki:1999}. Rates for $\beta$
delayed and spontaneous fission were adopted
from~\cite{Petermann.Langanke.ea:2012}. Our fission yields were taken
from the calculations of Ref.~\cite{Zinner:2007} which were derived
using the code ABLA\@. This approach also gives a consistent estimate
for the number of neutrons set free during the fission process.

Nuclear reactions change the energy balance of the environment. We
take this into account following
refs.~\cite{Freiburghaus.Rosswog.Thielemann:1999} by calculating, at
any time of the evolution, the change in abundances in the various
nuclei by solving the nuclear network. The related energy release is
mainly connected to $\beta$ decays where we assume that half of the
energy set free in the process is carried away by the
neutrino~\cite{Metzger.Arcones.ea:2010,Metzger.Martinez-Pinedo.ea:2010}.
In the next step, the energy release can be translated into a change
of entropy, from which we calculated a new temperature using the
equation of state of Ref.~\cite{Timmes.Arnett:1999} which considers
nucleons, nuclei, electrons, positrons and photons.

\section{Results}
\label{sec:results}

\subsection{Time evolution and energy generation}
\label{sec:energy-generation}

We have started our r-process simulations for 
all trajectories
at temperatures $T = 6$~GK assuming an initial
NSE composition. As a striking feature, we note the very large
neutron-to-seed ratios ($R_{n/s} \approx 400-2000$) which is a
prerequisite of r-process nucleosynthesis for nuclides beyond the
third peak reaching the region in the nuclear chart where nuclei decay
by fission. Furthermore, due to the extremely high neutron densities
involved, the path of the r process in NS mergers runs through nuclei
close to the dripline. For these nuclei, with their large $Q$ values,
$\beta$ decays are fast (order ms or faster).

Before we study the nucleosynthesis results of our simulations we like
to make some important remarks concerning the heating of the
astrophysical medium by nuclear reactions, mainly by $\beta$
decays. In Fig.~\ref{fig:reheat} we show the evolution of density,
temperature and the energy release due to nuclear reactions for 
all trajectories and the different mass models considered. 

The density and temperature profiles of the trajectories
are shown in Fig.~\ref{fig:reheat}.
The temperature evolution of the trajectories shows some characteristic pattern
which we will discuss in the following. 
For most of the trajectories, the temperature initially drops
to a minimum in the first few tens of ms and later rises to a maximum
during the r process. This
behavior can be understood from basic thermodynamics. From the first law of
thermodynamics the energy per nucleon released by nuclear reactions,
$\dot{q}$, is related to the change of the energy per nucleon,
$\varepsilon$ and nucleon density, $n=\rho/m_u$ ($m_u$ the atomic mass
unit), as:

\begin{equation}
  \label{eq:qdote}
  \dot{q} = \frac{d\varepsilon}{dt}-\frac{P}{n^2} \frac{dn}{dt} = c_V
  \frac{dT}{dt} + \left(\frac{d\varepsilon}{dn}-\frac{P}{n^2}\right)
  \frac{dn}{dt}. 
\end{equation}
where $c_V =d\varepsilon/dT$ is the specific heat per nucleon at
constant volume. The evolution of temperature reduces to:

\begin{equation}
  \label{eq:Tevol}
  \frac{dT}{dt} = \frac{1}{c_V} \left[\dot{q} -\frac{1}{\tau_n}
    \left(\frac{P}{n}-n\frac{d\varepsilon}{dn}\right)\right]
\end{equation}
where we have introduced the expansion timescale $\tau_n=1/\lambda_d$.
At early times when material expands from high densities,
the expansion time scale is rather small $\tau_n \approx$~1~ms  and the
second term in the bracket dominates and consequently the temperature
decreases as the material expands. However, as the expansion proceeds
and the temperature decreases there will be a moment at which both
terms on the right hand side become of the same magnitude. This will
correspond to a minimum in the temperature that can be estimated
assuming an equation of state dominated by nuclei (Boltzmann ideal
gas) and radiation (photons). In this case, we have

\begin{subequations}
  \label{eq:eos}
  \begin{eqnarray}
    \label{eq:enucleon}
    \varepsilon & = & \frac{3}{2}\frac{kT}{A} + \frac{a T^4}{n} \\
    \label{eq:Pressure}
    P & = & \frac{n kT}{A} + \frac{1}{3} a T^4 
  \end{eqnarray}
\end{subequations}
At initial times the average mass number $A$ is about 1, and it grows
as the r process proceeds. Substituting in equation~\eqref{eq:Tevol}
we have:

\begin{equation}
  \label{eq:Tevoldet}
  \frac{dT}{dt} = \frac{1}{c_V} \left[\dot{q} -\frac{1}{\tau_n}
    \left(\frac{4 a T^4}{3 n} +\frac{k T}{A}\right)\right]
\end{equation}
The minimum of temperature is reached when the right hand side of
the above equation is zero. Initially, when the density is relatively
high, the second term in the inner parentheses dominates and we have:

\begin{equation}
  \label{eq:Tmin}
  T_{\text{min}} = 0.05\ \mathrm{GK} \biggl(\frac{\dot{q}}{4\ \text{MeV
        s}^{-1}}\biggr) \biggl(\frac{\tau_n}{1\
      \mathrm{ms}}\biggr)\biggl(\frac{A}{1}\biggr) 
\end{equation}
for the minimal temperature assuming typical values for the other
quantities. We see that the minimum temperature is proportional to the
expansion timescale. This fact can be understood by noticing that
during the expansion the energy generation can only contribute
efficiently to increase the temperature during a period of time
$\tau_n$. Once the minimum value is reached the temperature starts to
rise favored by the fact that the timescale for the expansion
grows. However, as the temperature increases and the density decreases
the first term in the inner brackets of equation~\eqref{eq:Tevoldet}
increases. A maximum in temperature is reached once the equation of
state becomes dominated by radiation corresponding to a temperature of

\begin{eqnarray}
  \label{eq:Tmax}
  T_{\text{max}} & = & \left(\frac{3 n \dot{q} \tau_n}{4
      a}\right)^{1/4} \\ \nonumber
      = & 0.8 &\ \text{GK}\left[ \biggl(\frac{\rho}{10^5\ \text{g
           cm}^{-3}}\biggr) \biggl(\frac{\dot{q}}{4\ \text{MeV
        s}^{-1}}\biggr) \biggl(\frac{\tau_n}{10\
      \mathrm{ms}}\biggr)\right]^{1/4},
\end{eqnarray}
where we have used typical values for the density, energy generation
and expansion timescale. The above discussion clearly shows that the
behavior of temperature is determined by basic thermodynamics and it
is driven by the fact that at high densities the EoS is dominated by
the ideal gas component while at low densities radiation
dominates. The dominance of radiation implies a rather large specific
heat $c_V \approx 4 a T^3/n$ that reduces the efficiency at which the
energy generation can contribute to temperature increase. This means
that the maxima in temperature is rather flat as observed in
the simulations and depends on the expansion timescale at times $0.1\
\mathrm{s} \lesssim
t \lesssim 1$~s. 

For some trajectories matter is being heated by the shock 
to temperatures $T\gtrsim 6$~GK at rather low densities 
$\rho\sim 10^8$~g~cm$^{-3}$. For these cases, radiation quickly dominates the 
contribution to the total energy such that the temperature directly
approaches $T_{\rm max}$ (as a saddle point) without exhibiting a minimum.

We like to stress that, for a given trajectory and independent of the
early phase of the temperature evolution, the temperature during the r
process is rather determined by the density and the expansion
time-scale, as can be seen from Eq.~(\ref{eq:Tmax}).  We have
confirmed this in our calculations by varying the initial temperature
for the r-process simulations between $3$~GK and $10$~GK\@.  We find
that the early temperature evolution depends on the initial
conditions. However, in the later evolution, which is crucial for
r-process nucleosynthesis, the temperature profiles converge to rather
similar conditions caused by the balance of nuclear heating and
radiation dominance. One can say that the material loses the memory of
the particular thermodynamical conditions when it was ejected. This
aspect together with the large neutron-to-seed ratio, which makes the
nucleosynthesis insensitive to the initial composition, is fundamental
to achieve a robust r-process pattern in dynamical ejecta

After the r-process comes to an end with matter decaying back to
stability (this occurs after $\sim 1$-2~s for the slow ejecta and
$\lesssim 10$~s for the fast ejecta), the energy release due to nuclear
reactions drops and the temperature follows an adiabatic expansion
that for the radiation dominated conditions implies
$T \sim \rho^{1/3}$.  It is important to note that the temperature of
nearly $10^9$ K, which may be achieved during most of the r process
due to nuclear reheating, is sufficient to establish an
$(n,\gamma) \rightleftarrows (\gamma,n)$ equilibrium for a significant
part of the trajectories, shown in Figure~\ref{fig:reheat}.

The energy release shows quite distinct peak structures as function of
time. These differences are substantial as a function of mass model.
However, they lead to only minor differences in the evolution of
temperature as discussed above.  The peaks are related to matter being
accumulated and then breaking through the r-process waiting points at
the magic neutron numbers, where fission cycling, which we will
quantify below, induces repetitions of these processes.

\begin{figure*}[htb]
  \centering
  \includegraphics[width=0.82\linewidth]{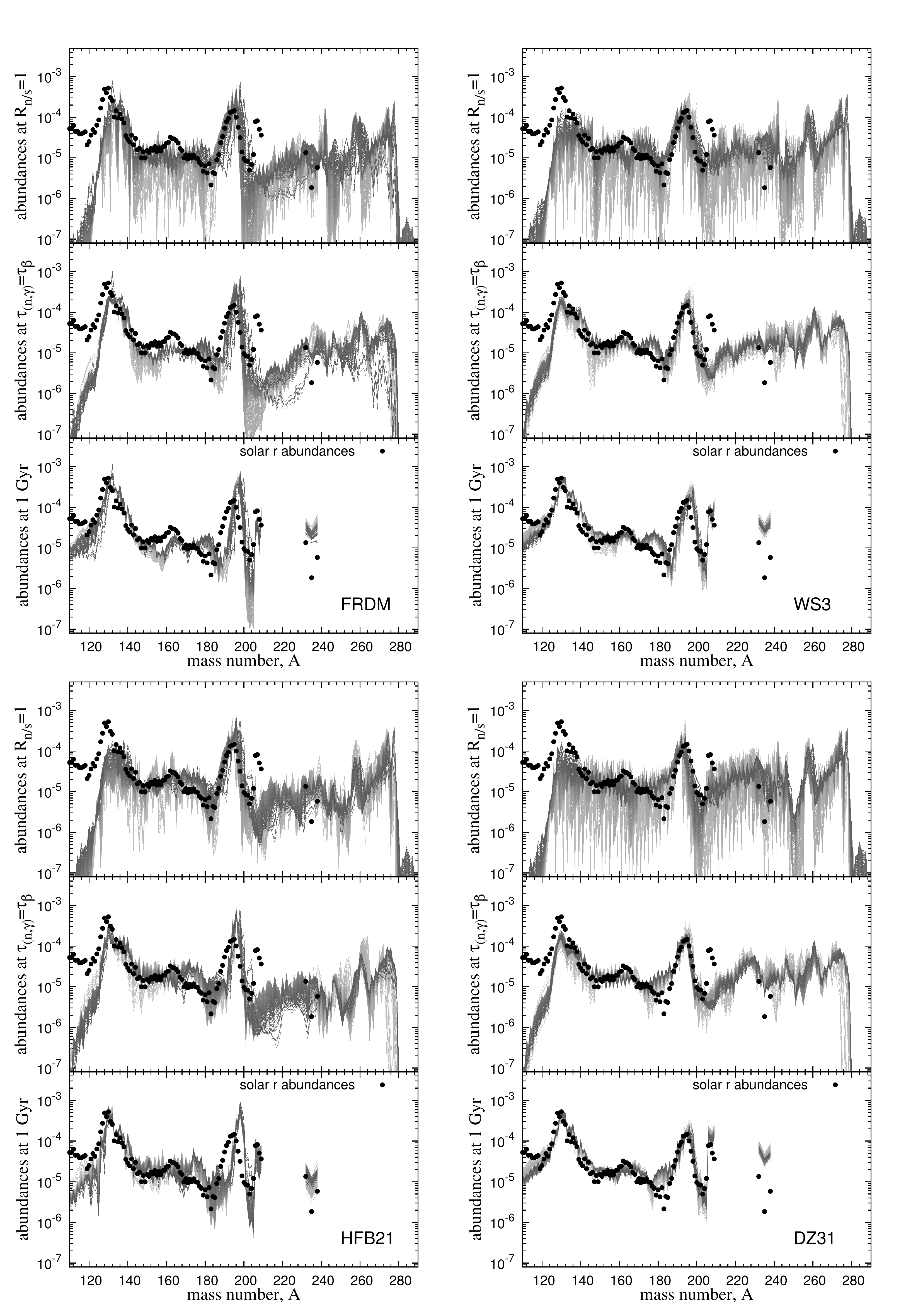}%
  \caption{(Color online) R-process abundances 
    of the slow ejecta for different mass models
    at different phases of the evolution. The upper
    panels show the abundances at $R_{n/s} = 1$. The middle panel at
    the time where the average timescales for beta-decay and neutron
    captures become identical. The lower panel shows the abundances at
    1~Gyr when most of the material has already decayed to the
    stability.\label{fig:abund-traj}}
\end{figure*}

\subsection{Evolution of the r-process abundances for the slow ejecta}
\label{sec:nucleosynthesis}

We have studied r-process nucleosynthesis for the set of 528 NS merger
trajectories and for 4 different mass models.  The calculated
r-process abundances show quite similar general patterns for all mass
models. In this subsection, we discuss the special features of
r-process nucleosynthesis of the so-called slow ejecta. The next
subsection then deals with r-process nucleosynthesis for the
fast-evolving ejecta.  For the slow ejecta we show the abundances at 3
different phases of the evolution: a) at freeze out, which we define
as the moment where $R_{n/s}=1$, b) the moment when the average
timescale for $\beta$ decays becomes equal to the average timescale
for neutron captures, c) the final abundance, calculated at a time of
1~Gyr.

Comparing the abundances at these 3 different phases for the slow
ejecta (shown in Fig.~\ref{fig:abund-traj}) we observe the following
important features. At freeze-out, the abundances show a strong
odd-even staggering which is washed out subsequently by continuing
neutron captures and $\beta$ decays towards stability. The magnitude
of the staggering is much larger for those trajectories that achieve
an $(n,\gamma)\rightleftarrows (\gamma,n)$ equilibrium, light gray
lines, compared with those which follow cold r-process conditions, dark
gray lines.  Strikingly, there is no abundance peak at $A \sim 130$,
in contrast to the third peak at $A \sim 195$ (the narrow peak at
around 136 is seen only in FRDM and will be discussed below). The
matter flow through the second peak is faster than through the third
peak. The origin of this difference is that even at the same neutron
separation energy, $Q_\beta$ values are larger at the $N=82$ nuclei on
the r-process path than for the $N=126$ nuclei making the respective
half-lives significantly shorter.  The second abundance peak is
produced mainly by fission yields from heavy nuclei around
$A \sim 280$ associated with the magic neutron number $N=184$. At
freeze-out, however, there is still a sufficient reservoir of free
neutrons from fission and subsequent $\beta$ decay and
photodissociation to support further neutron captures, also on nuclei
around $A \sim 130$. These neutron captures (note that
$\tau_\beta > \tau_{(n,\gamma)}$ at freeze out) shape the abundances
after freeze out significantly as can be seen when comparing the upper
and middle panels in Fig.~\ref{fig:abund-traj}. Importantly, we
observe that the second peak now forms (as transport from the fission
yields to heavier nuclei by neutron captures is reduced). Furthermore,
the strong abundance hole just above $A \sim 195$ is filling up due to
decays of heavier nuclei (mainly $\alpha$ decays of nuclei between
lead and thorium).  At the time of 1~Gyr, final abundances for mass
numbers $A > 120$ are virtually identical, for a given mass model, for
all slow ejecta as shown in the bottom panels. This points to an
extremely robust mechanism in shaping the final abundance pattern
nearly independent of the initial large variation of the astrophysical
condition which will be further discussed
in~\ref{sec:final-r-process}. Below in this subsection, we will focus
on the detailed discussion of the nuclear physics properties that give
rise to the variation seen in different nuclear mass models.

\begin{figure}[htb]
  \centering
 \includegraphics[width=\linewidth]{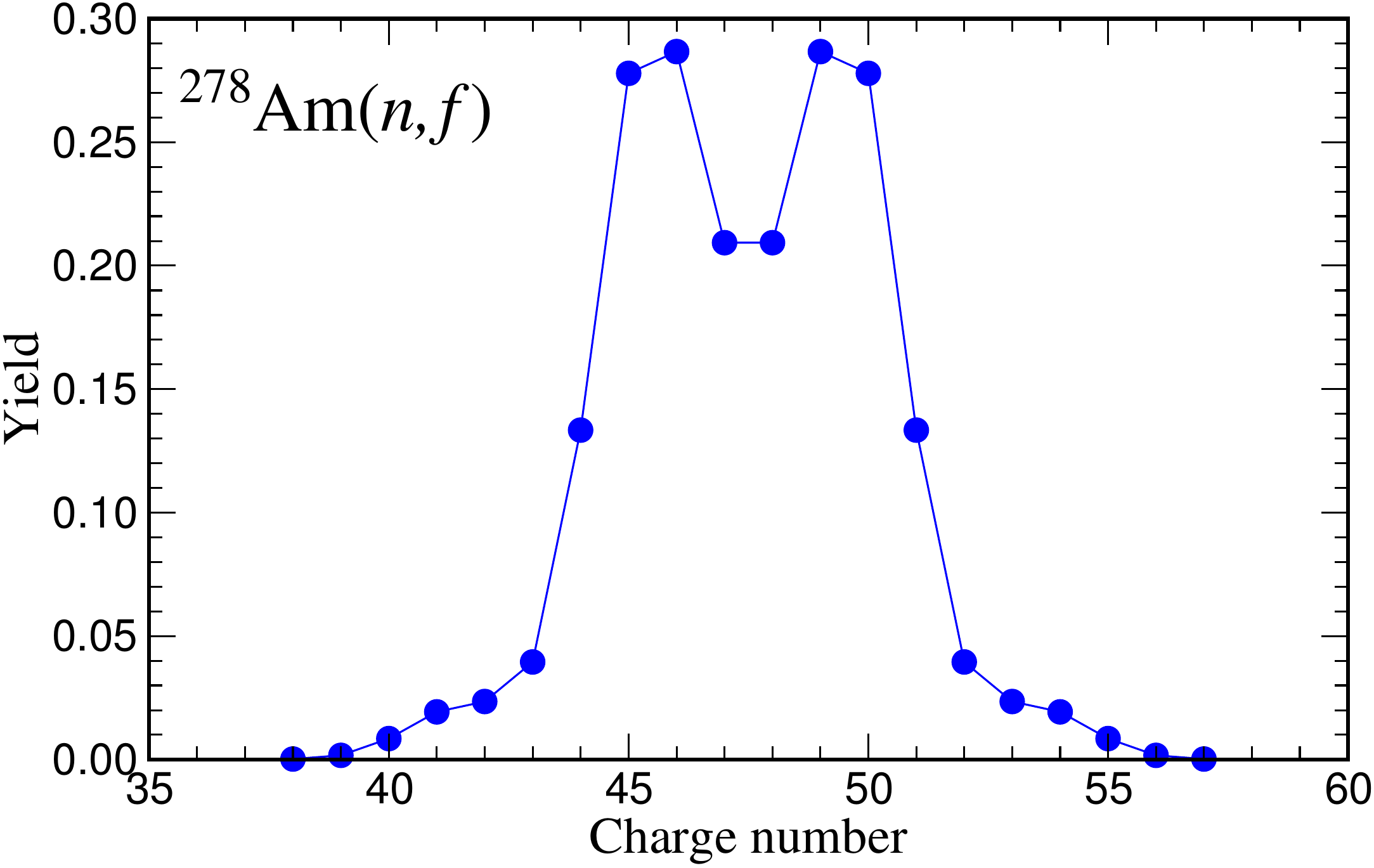}\\
 \includegraphics[width=\linewidth]{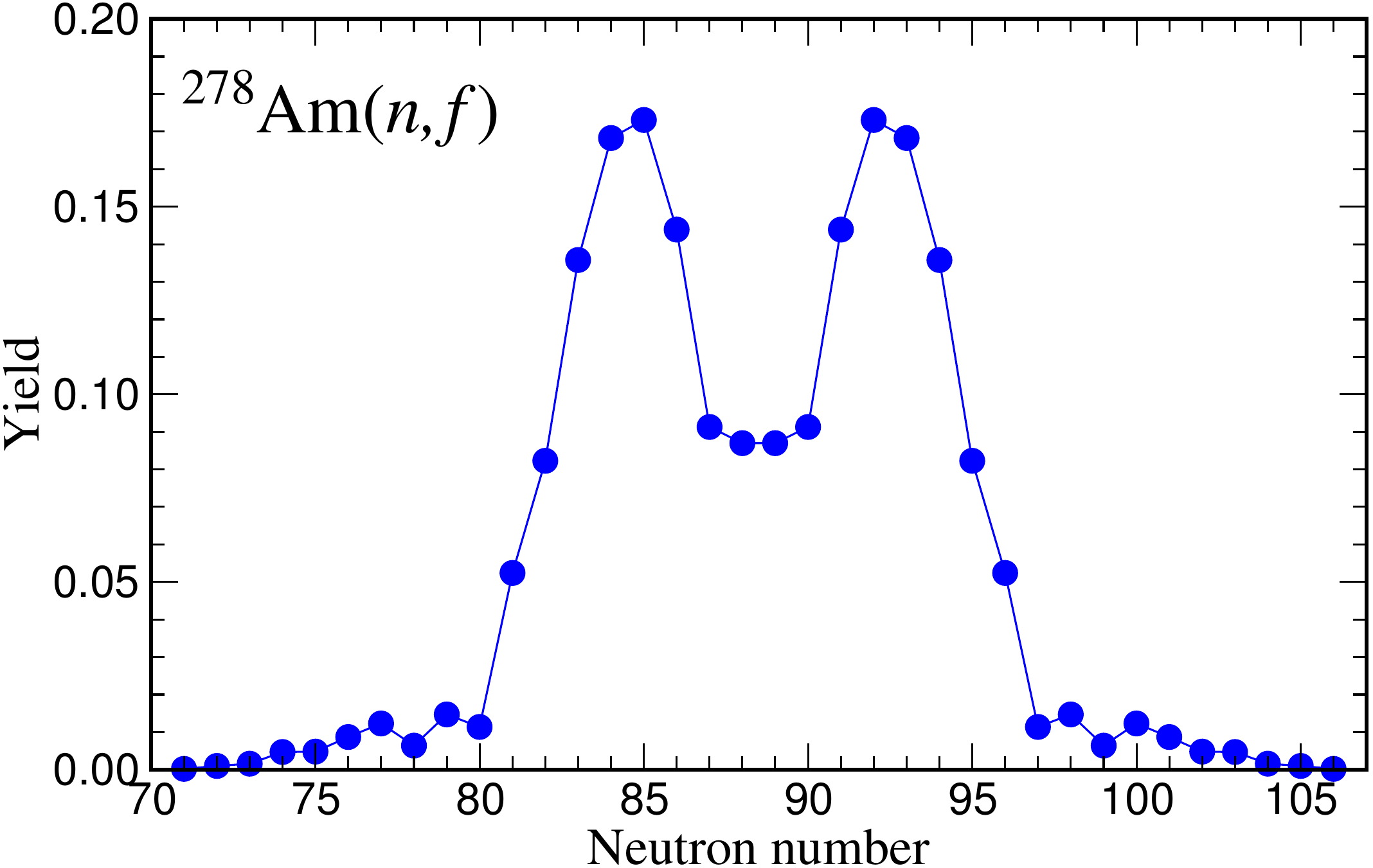}\\
 \includegraphics[width=\linewidth]{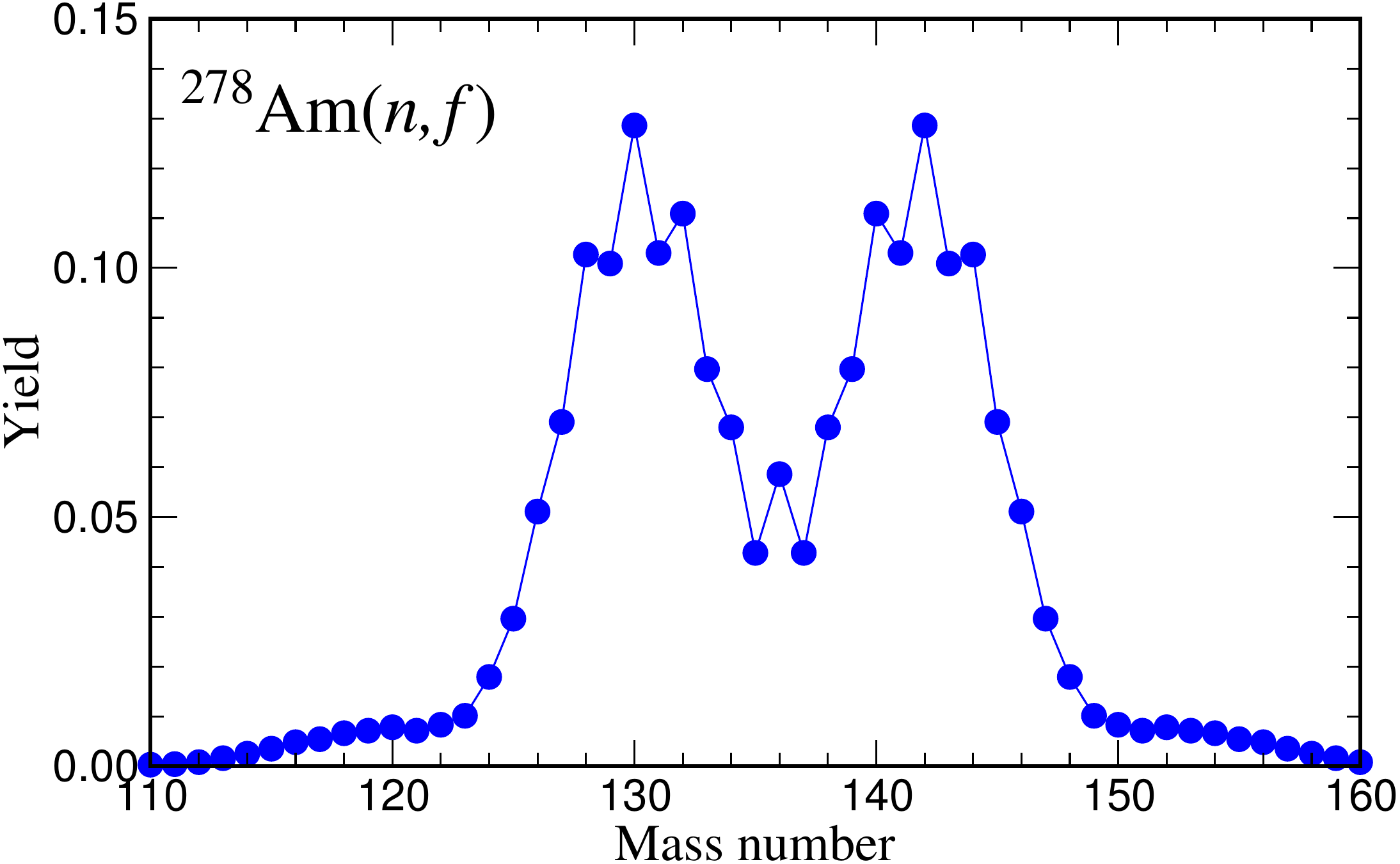}
  \caption{(Color online) Fission fragment distributions as functions
    of charge, neutron number, and mass number for neutron induced
    fission on $^{278}$Am as predicted by the ABLA
    code.\label{fig:fissionyields}}
\end{figure}

At freeze-out the nuclei with largest abundances in the transuranium
region are located around $A\sim280$ corresponding to the $N=184$
shell closure. The lightest nuclei for which neutron separation
energies are large enough to allow matter flow beyond $N=184$ is
$^{279}$Am. However, for this nucleus the fission barrier is so low
that neutron-induced fission on $^{278}$Am dominates over neutron
capture and the r-process cycles to medium mass nuclei rather than
producing heavier nuclei. Correspondingly, the nucleus with maximum
mass produced in our calculations is $^{278}$Pu that decays by
beta-emission subsequently followed by neutron-induced fission on
$^{278}$Am. The corresponding fragment distributions as a function of
mass, charge and neutron number are shown in
figure~\ref{fig:fissionyields}, indicating that one of the fragments
contributes to the second peak r-process abundances while the other is
produced around $A\sim 140$. Subsequent neutron captures and
beta-decays distribute the contribution of this second fragment to
higher masses. 

Fission is also an important source of free neutrons. The total number
of neutrons produced can be divided into two components. A prompt
component that consists of the neutrons evaporated mainly by the
highly excited fragments. For neutron-induced fission on $^{278}$Am,
ABLA predicts a prompt emission of 7 neutrons. In addition to the
prompt neutron emission, there is a delayed component that occurs
during the decay of the fragments to the instantaneous r-process
path. This delayed component depends on the location of the fragments
relative to the r-process path and the dominating reaction mechanism
(beta-decay or photodissociation). Both the r-process path and the
reaction mechanism sensitively depend on the astrophysical conditions,
particularly on the temperature. If the temperature is large enough,
photodissiociation is the dominating reaction mechanism. Under these
conditions the number of free neutrons produced is fully determined
knowing the charges $Z_1$ and $Z_2$ of the produced fragments and is
given by $n_{\text{prod}} = A_f - A_{\text{path}}(Z_1) -
A_{\text{path}}(Z_2)$, where $A_f$ is the mass number of the
fissioning nucleus and $A_{\text{path}}(Z)$ is the mass number of the
nucleus with charge $Z$ on the instantaneous r-process path. This
means that for the assumed astrophysical conditions, the total number of
free neutrons is completely determined by the charge fragment
distribution of the fission process. While photodissociation is the
dominating reaction mechanism at sufficiently high temperatures,
beta-delayed neutron emission becomes increasingly important with
decreasing temperature. In this case, the total number of neutrons
produced will depend on the prompt component predicted by the fission
model and the chain of beta-decays occurring during the decay to the
instantaneous r-process path. 

Once neutron captures are slower than $\beta$ decays, matter decays to
stability. In particular, the significant amount of matter above lead,
still existing in the middle panels of figure~\ref{fig:abund-traj},
decays to finally form the lead peak. At 1~Gyr only the long-lived
thorium and uranium isotopes survive. When comparing the time
evolution of the third peak (from the upper to the middle to the lower
panel of figure~\ref{fig:abund-traj}) one clearly notices a shift of
the third peak to slightly larger mass numbers in the FRDM mass model.
This is caused by neutron captures after freeze-out and confirms the
findings reported in~\cite{Eichler.Arcones.ea:2015}. Looking at the
final abundances computed for the mass models used in the present
study (see Fig.~\ref{fig:allmasses}) we observe that the shift in the
third r-process peak is present in the FRDM and HFB21 mass model but
absent in the WS3 and DZ31 mass models. To understand the reason for
this behavior it is important to qualitatively quantify which regions
of the nuclear chart are expected to be more relevant for the
evolution of the r process under very general assumptions.

The r process operates along a path of almost constant neutron
separation energy. The speed at which the r process proceeds from
lighter nuclei to heavier nuclei depends on the beta-decay
half-lives. Due to the increase in Coulomb energy the valley of
stability moves to more neutron rich nuclei with increasing charge
number. This means that for a line of constant neutron separation
energy the beta-decay Q-values reduce with increasing charge. As a
consequence, the beta-decay half-lives of r-process nuclei increase
with increasing mass number. On top of this global behavior, there are
local effects induced by the presence of neutron shell closures. In
particular, at neutron magic numbers $N=82$ and $N=126$ the r process
moves closer to stability to nuclei with longer beta-decay
half-lives. In the r-process path nuclei with $N\gtrsim 82$ and
$N\gtrsim 126$ have the longest half-lives. At freeze-out, it
corresponds to charge numbers $Z\approx 48$ and $Z\approx 70$,
respectively. The predicted half-lives of these
nuclei~\cite{Zhi.Caurier.ea:2013} are of the order of 100~ms which is
not much less than the total duration of the r process, namely around
1~s. Changing the time the r process expends in this long-lived nuclei
affects the whole dynamics of the r process and consequently impacts
the r-process abundances. This timescale can be affected by
modifications of individual beta-decay half-lives that so far are
based on relatively uncertain theoretical
approaches~\cite{Moeller.Pfeiffer.Kratz:2003,%
Marketin.Huther.Martinez-Pinedo:2015}. Alternatively, the effective
r-process timescale in the region can change if the r-process path
changes due to modifications of the underlying mass model. This is the
aspect explored in this work.

Different mass models differ substantially in their predictions in
regions where there is a sudden change in the intrinsic
deformation~\cite{Arcones.Martinez-Pinedo:2011}. This is particularly
the case around $N\sim 90$ and $N\sim 130$ where all mass models used
in the present work predict a transition from spherical to deformed
configurations. The particular relevance for the r process is the fact
that this transition can be associated with a sudden drop in the
neutron separations energies. This is the case for the FRDM mass model
as was already pointed out in Ref.~\cite{Meyer.Mathews.ea:1992} for
the Tellurium isotopes reaching $^{139}$Te. Recent mass measurements
for Tellurium
isotopes~\cite{Hakala.Dobaczewski.ea:2012,Van.Lascar.ea:2012} have
ruled out this sudden drop in neutron separation energies. However, at
present there is no data for lighter isotopes in the region where FRDM
also predicts very low neutron separation
energies~\cite{Arcones.Martinez-Pinedo:2011}. None of the other models
used in the present study show such a drastic reduction in neutron
separation energies around $N=90$. The most noticeable consequence is
the presence of a narrow peak around $A \sim$~136 at freeze-out (see
upper panel for FRDM mass model in Fig.~\ref{fig:abund-traj}). Due to
the accumulation of material in this region, the r process lasts
slightly longer using the FRDM mass model when compared with the other
models (see Fig.~\ref{fig:reheat} where the end of the r process is
associated to the drop in the heating rate). The peak becomes washed
out at later times due to continuous production of material in this
region by fission. However, neutron captures on the fission yields are
responsible for a flow of matter from the second r-process peak to
heavier nuclei. This flow operates in all used mass models except in
FRDM due to the fact that material is halted at $N\sim 90$.

The situation at the third peak is different as the fission yields
adopted here do not directly produce material in this region. The
third peak abundance is noticeably more sensitive to nuclear masses
that influence the neutron capture rates. It is not surprising that
the abundances of this peak show a larger variation. However, for two
of the mass models (FRDM, HFB21) the peak width is noticeably narrower
than observed, the peak height is overestimated, the position shifted
slightly to larger mass numbers and an abundance trough is predicted
just above the peak. This is related to different behavior of these
two models at the neutron number $N=130$, just above the magic number
$N=126$.  The FRDM and HFB21 mass models predict noticeably smaller
neutron separation energies than the Duflo-Zuker or the WS3 models in
this mass range.  For example the nuclei $^{199}$Yb, $^{198}$Tm and
$^{197}$Er (all with $N=129$) have neutron separation energies of
$S_n=0.52$ (0.85)~MeV, 0.62 (0.73)~MeV, and 0.26 (0.56)~MeV, in the
FRDM (HFB21) mass models~\footnote{We note that the low neutron
  separation energies remain in the latest version of the HFB type
  mass model, HFB27~\cite{Goriely.Chamel.Pearson:2013}}, respectively,
while they are 1.387 (1.463)~MeV, 1.479 (1.528)~MeV, and 0.908
(0.864)~MeV for the same nuclei in the DZ31 (WS3) models.  Thus these
nuclei act as (additional) obstacles in r-process simulations using
the FRDM and HFB21 mass models, even if the mass flow has overcome the
$N=126$ waiting points. As a result, the third peak in the abundance
distribution is shifted for these two mass models to higher mass
numbers as can be seen in Fig.~\ref{fig:abund-traj} caused mainly by
late-time neutron captures. Due to the larger neutron separation
energies, the $N=130$ nuclei do not act as obstacles in simulations
adopting the Duflo-Zuker or WS3 mass models. Relatedly the third peak
develops at $A \sim 195$, associated with the $N=126$ waiting points.

\begin{figure*}[htb]
  \centering
  \includegraphics[width=0.82\linewidth]{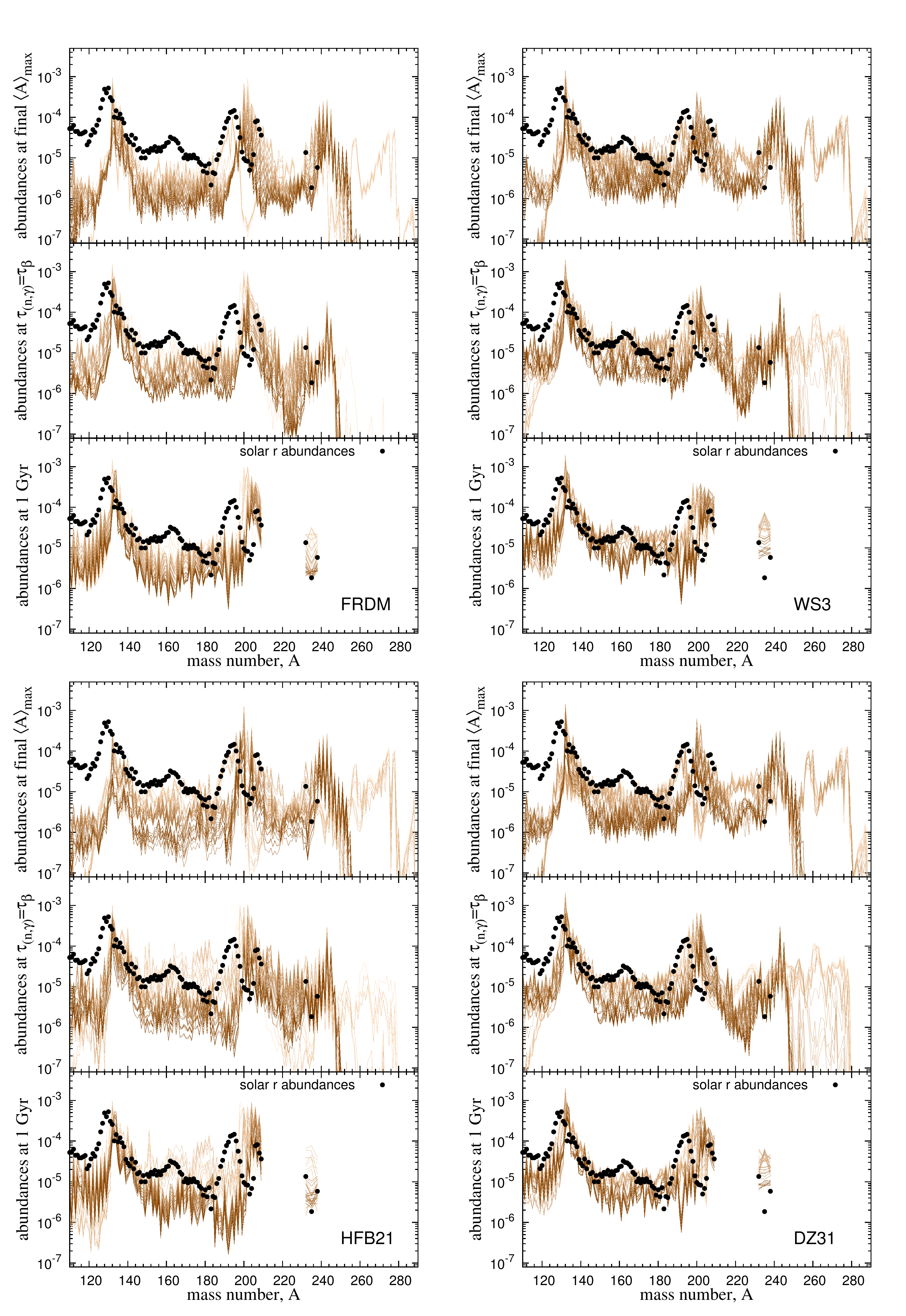}%
  \caption{(Color online) R-process abundances 
    of fast ejecta for different mass models
    at different phases of the evolution. The upper
    panels show the abundances at times before the ``last'' fission cycle,
    when the average mass number $\langle A\rangle$ reaches the final maximum. 
    The middle panel at
    the time where the average timescales for beta-decay and neutron
    captures become identical. The lower panel shows the abundances at
    1~Gyr when most of the material has already decayed to the
    stability.\label{fig:abund-traj-fast}}
\end{figure*}

\subsection{Evolution of the r-process abundances for fast ejecta}
\label{sec:nucleosynthesis-fast}

The characteristic of these trajectories is the very fast initial
expansion which results in an r process operating at much lower
densities than encountered for the `slow ejecta'. These low
densities translate into a slow neutron depletion rate, with the
important consequence, that a significant amount of neutrons remain at
the end of the r process.  (These neutrons will subsequently decay
contributing to the late-time r-process heating as discussed below in
Sec.~\ref{sec:impl-kilon-observ}.)  Hence, we define the freeze-out
for the fast ejecta at the moment when $\tau_{(n,\gamma)}=\tau_\beta$
rather than by $R_{n/s}=1$, as done for the slow ejecta.  Due to the
initially large neutron-to-seed ratios, all trajectories with fast
expansion undergo 1-3 fission cycles.

In Fig.~\ref{fig:abund-traj-fast} we show r-process abundances
obtained for the fast ejecta at three different phases of the
evolution and for the 4 different mass models.  The top panel, for
each mass model, shows the abundances before the ``last'' fission
cycle, when the average mass number $\langle A\rangle$ reaches the
final maximum. The middle panel exhibits the abundances for
$\tau_{(n,\gamma)}=\tau_\beta$ and the lower panel shows the final
r-process abundances at 1 Gyr.  Due to the slower neutron capture
rates, the r-process path for the fast ejecta runs noticeably closer
to the region of stability; i.e.\ through less neutron-rich nuclei,
than for the slow ejecta.  This has several consequences. First, the
position of third r-process peak, related to nuclei with magic neutron
number $N=126$, is shifted to larger mass numbers around $A \sim 200$.
This can be seen by comparing the r-process abundances, obtained for
the fast ejecta, at the times before the ``last'' fission cycle (top
panels of Fig.~\ref{fig:abund-traj-fast}), with those for the slow
ejecta at freeze-out, exhibited in Fig.~\ref{fig:abund-traj}.  Second,
under the conditions of fast expansion, and slow neutron captures, the
nuclei with magic neutron numbers $N=82$ are a noticeable obstacle for
the mass flow towards heavier nuclei. In particular, matter
accumulates, already before the last fission cycle, at the
double-magic nucleus $^{132}$Sn which has a relatively long beta-decay
half-life of $\approx 37$~s, producing a pronounced peak.  Third,
neutron captures are too inefficient to replenish the region of
$A \sim 280$ prior to the last fission cycle for most of the
trajectories.  As a consequence, the subsequent decay of these heavy
nuclei by fission contributes only rather modestly to the r-process
abundances around the second peak at $A \sim 130$, as can be seen in
the lower panels of Fig.~\ref{fig:abund-traj-fast}.  However, the
decay of matter beyond the third peak, after freeze-out, fills up the
abundances around lead.

In contrast to the slow ejecta, the fast ejecta exhibit a large
spread in the final abundances observed between the different
trajectories. This points to a very strong sensitivity to details of
the astrophysical conditions and to the nuclear properties, if neutron
captures are slow during the r process. In fact, the fast ejecta,
encountered in our NS merger scenario, resemble a nucleosynthesis
process somewhat between r process and s process, producing abundance
peak structures shifted noticeable to larger mass numbers than
observed in the solar r-process abundance distribution.

\subsection{Robust r-process abundances}
\label{sec:final-r-process}

\begin{figure*}[htb]
  \centering
  \includegraphics[width=\linewidth]{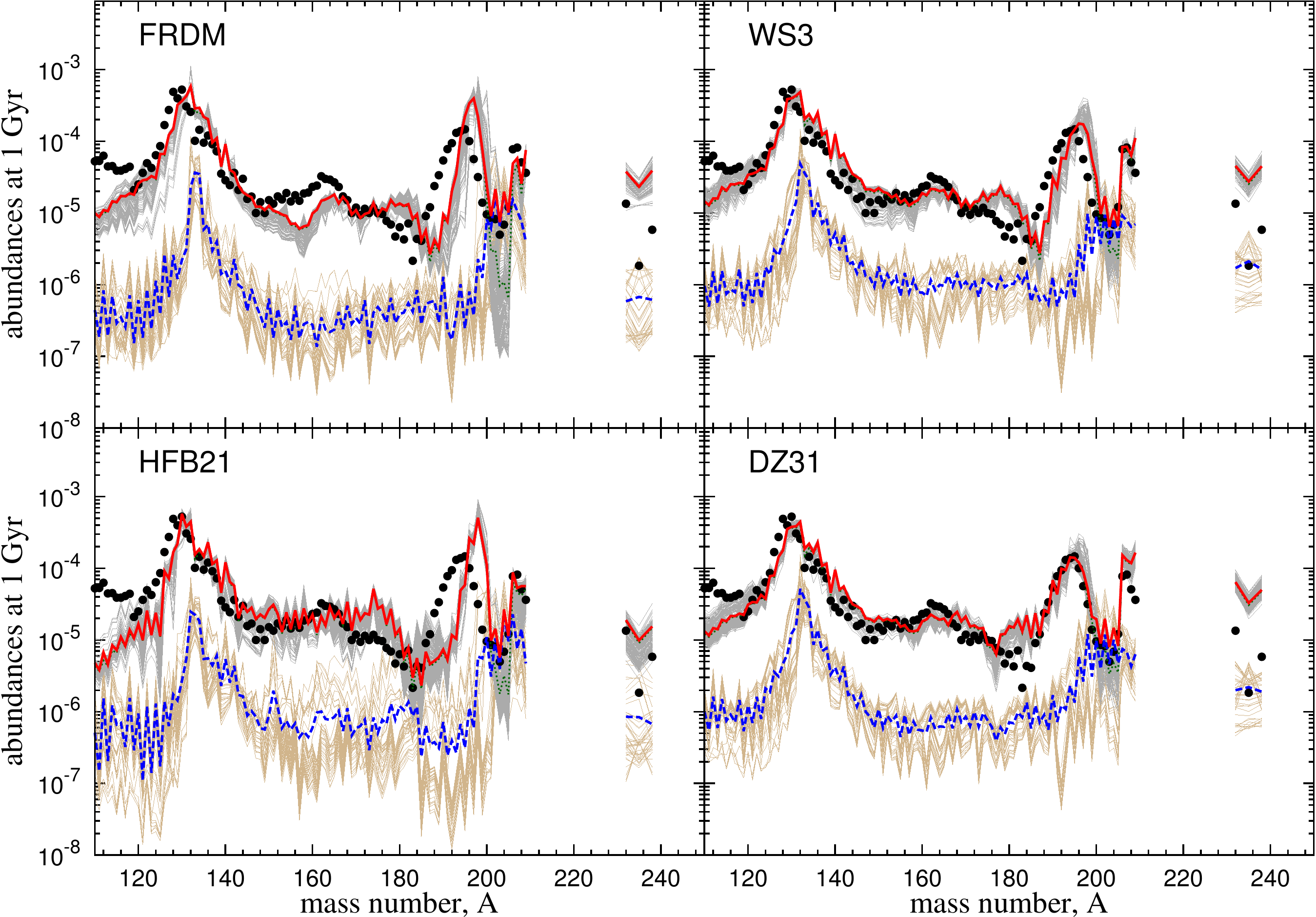}
  \caption{(Color online) Final r-process abundances at a time of
    1~Gyr for the different mass models and all trajectories used in
    the calculations. The grey (brown) curves correspond to the
    abundances of the trajectories of the slow (fast) ejecta shown
    previously in the bottom panels of Fig.~\ref{fig:abund-traj}
    and~\ref{fig:abund-traj-fast} but without the color gradient. The
    mass-averaged abundances for all trajectories (red curves), the
    slow ejecta (green curves), and the fast ejecta (blue curves)
    are also shown. The abundances for the slow and fast
    trajectories and their averages have been scaled by the value of
    their fractional contribution to the total
    ejecta.\label{fig:finalabund}}
\end{figure*}

\begin{figure}[htb]
  \centering
  \includegraphics[width=\linewidth]{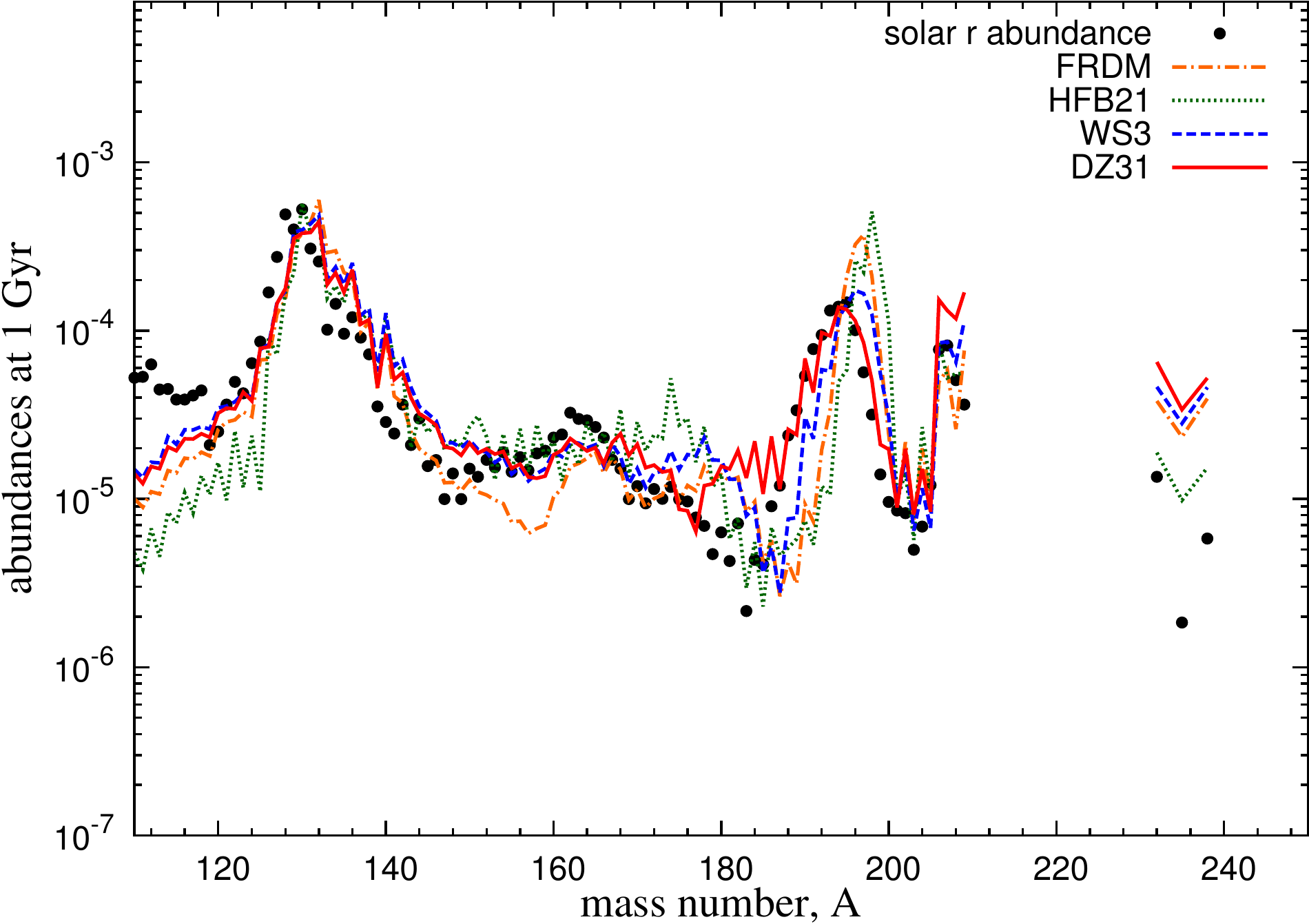}
  \caption{(Color online) Final mass-integrated abundances
   for all trajectories at a time of 1~Gyr for all
    mass models considered in this work.\label{fig:allmasses}}
\end{figure}

Fig.~\ref{fig:finalabund} shows the final abundances at times of 1~Gyr
for all individual NS merger trajectories and for all the mass models.
Additionally, the figure exhibits mass-averaged abundances for all
trajectories (red curves), for the slow trajectories (green curves),
and for the fast ejecta (blue curves), respectively. To better
visualize the contribution of slow and fast ejecta to the total
ejected mass we have multiplied the slow and fast trajectories and
their averages by the fractional contribution of each ejecta, i.e.\
$\sim 0.9$ for the slow and $\sim 0.1$ for the fast ejecta

As already stressed above, the most striking feature of our
calculations is the fact that the final abundances for mass numbers
$A > 120$ are virtually identical, for a given mass model, for all
the slow ejecta, while they vary noticeably for the fast ejecta.
Furthermore, the total mass-averaged abundances show the same
pattern as those for the slow ejecta, as these constitute the
dominating part of the ejected mass.  We hence conclude that dynamical
ejecta of NS mergers show a robust r-process pattern, as already
concluded in
refs.~\cite{Goriely.Bauswein.Janka:2011,Korobkin.Rosswog.ea:2012,%
  Bauswein.Goriely.Janka:2013}, provided that $Y_e$ in the ejecta
remains low, see
refs~\cite{Wanajo.Sekiguchi.ea:2014,Sekiguchi.Kiuchi.ea:2015,Goriely.Bauswein.ea:2015},
as expected in dynamical ejecta of NS-black hole and very asymmetric
NS-NS mergers. We will now discuss the origin of the robustness. 

Thanks to the large neutron-to-seed ratio found in NS merger
conditions the r process runs through 1--4 fission cycles, where
trajectories with larger initial neutron-to-seed ratio obviously
support more cycles. This so-called fission cycling has been suggested
to be responsible of producing r-process abundances that are almost
independent of the astrophysical
conditions. Ref.~\cite{Beun.Mclaughlin.ea:2008} has suggested that
fission cycling contributes to produce a steady $\beta$ flow
equilibrium in which the abundances for each isotopic chain are
proportional to the beta-decay half-lives. Steady $\beta$ flow
equilibrium is in fact achieved in NS merger before the r-process
freeze-out as the duration of the r process is longer than the
individual beta-decay half-lives. Furthermore, it can be achieved both
for hot and cold r-process
conditions~\cite{Arcones.Martinez-Pinedo:2011}. The upper panels of
Fig.~\ref{fig:abund-traj} show substantial differences in the
r-process abundances for the different trajectories that nevertheless
converge to a robust abundance pattern at the end of the
calculations. This suggests that the mechanism responsible for
producing a robust r-process pattern operates after r-process
freeze-out and it is independent of the number of fission cycles.

We find that the main requirement to achieve a robust r-process
pattern is that the amount of material accumulated at freeze-out in
the fissioning region, $A\gtrsim 250$ is much larger than the one
present in the region below the 3rd r-process peak. 
For the slow ejecta, this is guaranteed
by the fact that the beta-decay half-lives grow with increasing mass
number and by the presence of a neutron shell closure around
$N=184$. Both effects are responsible of producing a peak in the
freeze-out r-process abundances around $A\sim 280$ (see upper panels
Fig.~\ref{fig:abund-traj}). The material in this peak decays by
fission contributing to the abundances around the 2nd
r-process peak (see figure~\ref{fig:fissionyields}) and producing a
final robust r-process pattern. 
For the fast ejecta, the lack of material accumulated in the $A>250$
region, due to the much slower neutron capture rates at later stage of
the r-process, results in reduced impact of fission yields 
on the final distributions 
and in a much larger spread of the final  abundance distributions. 
However, these trajectories contribute only mildly
to the final mass-integrated abundances except for the region
around $A=200$.

The general features of
this pattern is also independent of the mass models as it is mainly
determined by fission yields. This is demonstrated in
Fig.~\ref{fig:allmasses} where we compare
the final mass-integrated abundances (at 1~Gyr) for four different mass models (FRDM,
WS3, HFB21, DZ31).
Although there are
specific differences originating in the dependence of neutron captures
on the underlying mass model, all the calculations reproduce the
second and third r-process peaks reasonably well. We mention again
that the peaks have different origins in our simulations: the peak
around $A \sim 130$ arises from fission yields, while the peak at $A
\sim 195$ reflects the $N=126$ waiting points in the matter flow
towards heavy nuclei.  It is also satisfying to observe that the lead
peak around $A \sim 208$ agrees reasonably well with the solar
abundances. This peak is mainly produced by $\alpha$ decay of heavier
nuclei. Finally also the abundances of the long-lived isotopes
$^{232}$Th and $^{238}$U, which are the final product of some matter
with charge numbers $Z \approx 90-96$, is reproduced reasonably well.

In details, there are shortcomings of our various calculations when
compared to the solar abundances. While the height and the width of
the $A \sim 130$ peak are well described by all models, the peak
position - mainly due to late-time neutron captures - is slightly
shifted to larger mass numbers than observed. The similarity of all
models in the description of this peak is related to the fact that we
use the same fission yield distributions in all studies. Furthermore
our description is noticeably improved compared with those of
Refs.~\cite{Goriely.Bauswein.Janka:2011,Korobkin.Rosswog.ea:2012,%
  Bauswein.Goriely.Janka:2013} due to the different fission barriers
and yields used in the present work. 

Goriely~\emph{et al.}~\cite{Goriely.Sida.ea:2013} have recently
presented a new fission fragment distribution model that predicts
substantially different fission yields for r-process nuclei to those
used in the present work. Given the important role that fission yields
play in determining the r-process abundances, it is rather important
to further study the sensitivity of the r-process abundances to the
fission yields, including the recently developed GEF
model~\cite{Schmidt.Jurado:2010,Schmidt.Jurado:2011,*Schmidt.Jurado:2011e,Schmidt.Jurado.Amouroux:2014},
and explore experimental possibilities to constrain them. This goes
beyond the goals of the present work.

\begin{figure}[htb]
  \centering
  \includegraphics[width=\linewidth]{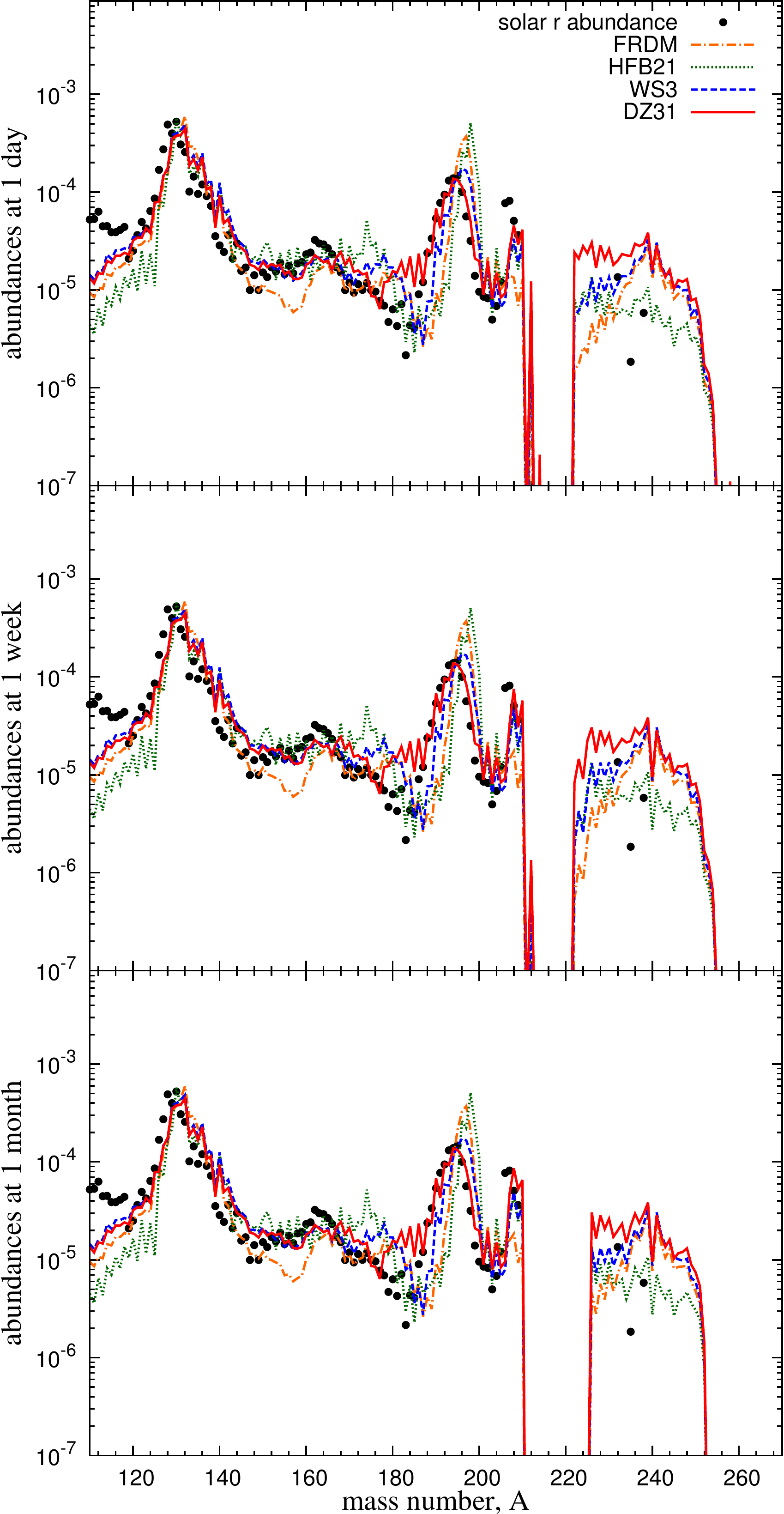}
  \caption{(Color online) Mass- r-process abundances on
    timescales of a day (upper panel), a week (middle panel) and a
    month (lower panel) after the r-process event for the different
    mass models considered in the present study. \label{fig:kilo}}
\end{figure}

\subsection{Implications for kilonova observations}
\label{sec:impl-kilon-observ}

One of the most interesting consequences of r-process nucleosynthesis
in neutron star mergers is that the large amount of material ejected
can produce an electromagnetic transient powered by the radioactive
decay of r-process
nuclei~\cite{Li.Paczynski:1998,Kulkarni:2005,Goriely.Bauswein.Janka:2011,Metzger.Martinez-Pinedo.ea:2010,Roberts.Kasen.ea:2011,%
  Bauswein.Goriely.Janka:2013}. This transient is commonly denoted as
kilonova and may have been recently observed associated with the short
$\gamma$-ray burst
GRB130603B~\cite{Tanvir.Levan.ea:2013,Berger.Fong.Chornock:2013}.

Modeling kilonova light curves requires the knowledge of the r-process
heating rates at timescales between hours and weeks and identify those
nuclei that predominantly contribute to the heating rate. This aspect
will be explored in further work. In addition, it is also important to
know the optical opacities for r-process material as they determine
the timescale for photons to escape from the inner opaque region of
the ejecta. It has been suggested that a major source of opacity is
due to the presence of lanthanides ($Z=57$--71, $A\approx
139$--176)~\cite{Kasen.Badnell.Barnes:2013,Barnes.Kasen:2013,Tanaka.Hotokezaka:2013}
in the r-process ejecta. Due to their large photon opacities the
lanthanides delay the light curve luminosity peak to timescales of a
week. The peak frequency is also shifted to the red. An additional
source of opacity could be due to the presence of actinides in the
r-process ejecta. The r process in NS mergers ejecta is known to
produce the actinides U and Th (see for example
Fig.~\ref{fig:allmasses}). However, at timescales of weeks relevant to
kilonova observations the abundance of actinides ($Z=89$--103,
$A\approx 225$--260) can be much larger since many of the nuclei
involved have long decay half-lives. This is confirmed in
Fig.~\ref{fig:kilo} that shows the mass-integrated 
r-process abundances at timescales of one day, one week and one month. 
As can be seen from Fig.~\ref{fig:abund-traj}, we find a
rather small dependence of the actinide abundances on the
astrophysical conditions. There is a much larger dependence on the
nuclear physics input (see Figs.~\ref{fig:abund-traj}
and~\ref{fig:kilo}). This aspect requires further
exploration. However, our results already show a strong correlation
between the behavior of neutron separation energies around $N\sim 130$
and the amount of actinides produced. Mass models that predict low
neutron separation energies tend to accumulate more material in the
third r-process peak and predict lower abundances for actinides. 

On top of the light curves due to the r-process heating, neutrons
that survived after the r process in the fast ejecta will decay and may
provide another electromagnetic probe for the neutron star mergers
that peaks hours after merger~\cite{Metzger.Bauswein.ea:2015}.  One
important quantity in determining the luminosity and the frequency
peak of this neutron-decay powered light curves is the amount of free
neutrons after the r process. We find in our calculations that this
amount is rather sensitive to the nuclear mass model adopted. For the
four different mass models we have explored, the total mass of free
neutrons at $\sim 20$ second post-merger is a sizable fraction of the
total mass of the fast ejecta and amounts to $5.39\times 10^{-5}$,
$4.42\times 10^{-5}$, $2.86\times 10^{-5}$, and
$2.68\times 10^{-5}$~M$_\odot$ for the FRDM, HFB21, WS3, and DZ31 mass
model, respectively. This sensitivity of the mass model is due to the
same reason discussed above that the low neutron separation energies
predicted by FRDM and HFB21 give rise to the smaller neutron capture
rates around the third peak such that there are free neutrons left at
the end of the r process, when compared to the WS3 and DZ31 mass
model. Moreover, we note that the opacity due to the r-process nuclei
for the fast ejecta may be somewhat different from the slow ejecta
which in turn may also affect the detailed modeling of the kilonova
light curves.

\section{Conclusions}
\label{sec:conclusions}

In summary, we have performed r-process simulations for matter ejected
in neutron star merger calculations.  Due to the extreme
neutron-to-seed ratios achieved, which favor mass transport to the
region of fissioning nuclei, fission rates and yields are crucial
ingredients in such calculations. Adopting these quantities from the
ABLA code which has been adjusted to reproduce fission and
fragmentation data and treats the release of neutrons during the
fission process explicitly, we have been able to reproduce the main
features of the r-process abundances (the second and third peaks, the
rare-earth peak and the lead peak) reasonably well. Importantly, we
found that these features do not depend sensitively on the
astrophysical conditions for the majority of the ejecta (but see
Ref.~\cite{Wanajo.Sekiguchi.ea:2014,Sekiguchi.Kiuchi.ea:2015,Goriely.Bauswein.ea:2015}
for a possible increase of the initial electron fraction which may
have an impact on the abundance patterns). We have also shown that
these features do not depend in general on the nuclear mass model
used. We noticed, however, modest differences in the position of the
third peak and in abundance distribution just above this peak around
$A \sim 205$. Here the FRDM and HFB21 mass models predict noticeably
smaller neutron separation energies for r-process nuclei with $N=130$
than the other mass models used in our studies. These small separation
energies make the $N=130$ nuclei obstacles in the r-process path
resulting in the peak shift and a pronounced abundance trough at
$A \sim 205$, if compared to the solar r-process
abundances. Experimental work is needed to resolve these different
mass predictions for the $N=130$ nuclei.

Our simulations support the hypothesis that the r process in dynamical
ejecta from neutron star mergers yield rather robust abundance
distribution (provided $Y_e$ of the ejecta is sufficiently low) in
good agreement with the observed solar distribution for nuclei with
$A\gtrsim 120$. We have shown that a requirement to achieve such a
robust pattern is that at freeze-out the amount of material
accumulated in the fissioning region ($A\gtrsim 250$) is much larger
than the material located in the second r-process peak and above
($A\approx 120$--180). To achieve these astrophysical conditions, a
sufficiently large neutron-to-seed ratio is required, which, together
with the fact that beta-decay half-lives along the r-process path grow
with increasing mass number, guarantees the pile up of material in the
fissioning region. The decay of this material by fission produces a
robust r-process pattern in the region $A\approx 120$-180 that,
however, depends on the used fission yields (see
ref.~\cite{Goriely.Martinez-Pinedo:2015}). This pattern is slightly 
modified by neutron captures during the decay back to stability, which
also introduces a small dependence of the abundance patterns on the
astrophysical conditions.  In this scenario, larger
neutron-to-seed ratios only increase the amount of fission cycles
without modifying the final r-process abundance distribution. These,
however, might be sensitive to the detailed shell structure in the
mass region of fissioning nuclei and, in particular, to the strength
of the $N=184$ shell closure.

Finally, in our simulation, $\sim 10\%$ of the ejecta expand so fast
that the resulting r-process operates at much lower neutron number
densities. Consequently, the nucleosynthesis path is closer to
stability with significantly less amount of fissioning nuclei present
at the time of freeze-out. Hence, the fission yields do not contribute
much to the final abundances at the second r-process peak and the
second and third r-process peaks shift to larger mass
numbers. Furthermore, there exists a large variation in the final
abundances between different trajectories, in clear contrast to the
slow ejecta.  In addition, a significant amount of neutrons is left
at the end of the r process. This amount, however, depends sensitively
on the nuclear mass model input.

The observation of electromagnetic signals of the r process is an
intriguing possibility. In this context our simulations imply that for
most of the mass models used the amount of actinides present at
timescales of several weeks is similar or larger than that of
lanthanides. Consequently, actinides may dominate the photon opacities
for kilonova light curves.  The differences in abundance pattern and,
also, the amount of free neutrons present at the end of the r process
for fast ejecta might be relevant for detailed modelling of these
light curves.

Current constraints on the NS merger rate are compatible with
NS-NS and/or NS-black hole mergers being the dominant source of 
r-process elements in our galaxy~\cite{Bauswein.Ardevol.ea:2014}. 
It is not excluded that the NS merger scenario is responsible
also for the solar-like r-process patterns observed in old,
metal-poor stars. This requires, however, that the
frequency of neutron star mergers is sufficiently large during
the early evolution of the Milky Way, which is a topic currently
being explored in the context of dynamical galactic chemical 
evolution models~\cite{Voort.Quataert.ea:2015,Shen.Cooke.ea:2015}.

Further extensions of the present work, need to address the impact of
variations of fission rates and beta-decay rates. In this study, we
have limited the variation of nuclear masses to nuclei with $Z<83$ as
for heavier nuclei fission rates have not been computed consistently
with the mass models used. Given the fundamental role played by
fission in dynamical ejecta of neutron star mergers, it is very
important to consistently determine both nuclear masses and fission
relevant quantities. These include both fission barriers and
collective inertial masses. Recent global
calculations~\cite{Goriely.Hilaire.ea:2009,Moeller.Sierk.ea:2009,%
Erler.Langanke.ea:2012}
have mainly focussed on the description of fission barriers neglecting
the important role of collective masses that are normally described
using a simple phenomenological prescription~\cite{Goriely:2015} in
r-process applications. Furthermore, it is important to address the
impact of dynamical versus static descriptions of fission
observables~\cite{Sadhukhan.Mazurek.ea:2013,%
Giuliani.Robledo.Rodriguez-Guzman:2014}

Beta-decay half-lives have not been consistently computed with the
mass models explored in this work. Ideally, one should use the same
model applied for the calculation of nuclear masses to determine the
Gamow-Teller and first-forbidden strengths that determine the
beta-decay half-lives of r-process nuclei. Consistency between the
calculations of masses and beta-decay half-lives can in principle be
achieved in mean-field approaches by performing quasiparticle random
phase approximation calculations where the residual interaction is
derived from the same density functional used for the calculation of
nuclear masses~\cite{Engel.Bender.ea:1999,%
Marketin.Huther.Martinez-Pinedo:2015}. However, this approach does
not uniquely determines the computed beta-decay half-lives as they are
quite sensitive to the choice of proton-neutron $T=0$
pairing~\cite{Engel.Bender.ea:1999,Niu.Niu.ea:2013} while this
interaction channel barely affects the masses of neutron-rich
r-process nuclei.


\begin{acknowledgments}
Helpful conversations with Marius Eichler and Friedel Thielemann about
the role of neutrons produced by fission in r-process nucleosynthesis
are acknowledged.  This work was supported by the Deutsche
Forschungsgemeinschaft through contract SFB~634, the Helmholtz
International Center for FAIR within the framework of the LOEWE
program launched by the state of Hesse, the Helmholtz Association
through the Nuclear Astrophysics Virtual Institute (VH-VI-417) and the
ExtreMe Matter Institute EMMI in the framework of the Helmholtz
Alliance HA216/EMMI\@.  A.B. is a Marie Curie Intra-European Fellow
within the 7th European Community Framework Programme (IEF
331873). This work was supported by the Deutsche
Forschungsgemeinschaft through Sonderforschungsbereich Transregio 7
``Gravitational Wave Astronomy'', and the Cluster of Excellence EXC
153 ``Origin and Structure of the Universe''.

\end{acknowledgments}

%

\end{document}